\documentclass[aps,prl,twocolumn,showpacs]{revtex4}
\usepackage{graphicx}
\usepackage{dcolumn}
\usepackage{amsmath}
\usepackage{amssymb}
\usepackage{amsmath,verbatim,bm}
\def\rv{{\bf r}}

\def\uv{{\bf u}}

\begin{document}
\title{Quantum Stress Focusing in Descriptive Chemistry} 
\author{Jianmin Tao$^{1,2}$, Giovanni Vignale$^2$, and I. V. Tokatly$^{3,4}$}
\affiliation{$^1$Theoretical Division and CNLS, Los Alamos National Laboratory, Los Alamos, New Mexico 87545 \\
$^2$Department of Physics, University of Missouri-Columbia, Columbia, Missouri 65211 \\
$^3$European Theoretical Spectroscopy Facility (ETSF), Dpto. Fisica Materiales,
 Universidad del Pais Vasco, 20018 Donostia, Spain\\
$^4$Moscow Institute of Electronic Technology, Zelenograd, 124498 Russia}
\date{\today}
\begin{abstract}
We show that several important concepts of descriptive chemistry, such as 
atomic shells, bonding electron pairs and lone electron pairs, may be 
described in terms of {\it quantum stress focusing}, i.e. the spontaneous 
formation of high-pressure regions in an electron gas. This  description 
subsumes previous mathematical constructions, such as the 
Laplacian of the density and the electron localization function, and 
provides a new tool for visualizing chemical structure.  We also show 
that the full stress tensor, defined as the derivative of the energy 
with respect to a local deformation, 
can be easily calculated from  density functional theory.  
\end{abstract}


\maketitle
Atomic shell structure, electron pair domains, $\pi$-electron subsystems, 
etc., are common concepts 
in descriptive chemistry and play a significant role in modern electronic 
structure theory. These concepts help us to visualize the bonding between 
atoms in terms of small groups of 
localized electrons (e.g. two electrons of opposite spin in a simple 
covalent bond) and therefore play an important role in predicting new 
molecular structures and in describing structural changes due to chemical 
reactions. 

A precise quantitative description of small groups of 
localized electrons has been sought for a long time~\cite{bader,gillespie}, 
but none of the solutions proposed so far is completely satisfactory.  
%\item 
The natural candidate -- the electronic density $n(\rv)$ -- 
has a clear physical meaning, but 
fails to reveal quantum mechanical features such as atomic shell 
structure~\cite{pp03} or the localization of electron pairs of opposite 
spin in a covalent bond.
%\item 
The Laplacian $\nabla^2 n$ of the density~\cite{bader,bader84} provides 
a better way to visualize molecular geometry, but lacks a clear physical 
significance and fails~\cite{be90} to reveal the atomic shell structure 
of heavy atoms.  
%\item  
Recently, a very useful indicator of electron localization has been 
constructed~\cite{be90} from the curvature of the (spherically averaged) 
conditional pair probability function $P_{\sigma\sigma}(\rv,\rv')$ 
(the probability of finding an electron of spin $\sigma$ at $\rv$ given that
there is another electron of the same spin orientation at $\rv'$) 
evaluated at $\rv=\rv'$.  In the simplest approximation
this curvature is proportional to
\begin{eqnarray}\label{belf}
D_\sigma (\rv)= \tau_\sigma - \frac{|\nabla n_\sigma|^2}{8n_\sigma},
\end{eqnarray}
where
%\begin{eqnarray}\label{tkin}
$\tau_{\sigma}(\rv) = \sum_{l}\frac{1}{2}
|\nabla\psi_{ l \sigma}(\rv)|^2$
%\end{eqnarray}
is  the non-interacting kinetic energy density of $\sigma$-spin electrons 
(atomic units $m = \hbar = e^2 = 1$ are used throughout), 
$\psi_{ l \sigma}(\rv)$ are the occupied Kohn-Sham orbitals of spin $\sigma$, 
and $n_\sigma$ is the $\sigma$-spin electron density 
($n = n_\uparrow + n_\downarrow$).  
From $D_\sigma$ one constructs the {\it electron localization function} 
(ELF)\cite{be90}
\begin{equation}\label{ELF}\eta_\sigma(\rv) = 
\frac{1}{1 + (D_\sigma/\tau_\sigma^{\rm TF})^2}, 
\end{equation}
where $\tau_\sigma^{\rm TF} = (3/10)(6\pi^2)^{2/3}n_\sigma^{5/3}$ is the 
Thomas-Fermi kinetic energy density of $\sigma$-spin electrons.  
The ELF provides excellent visualization of atomic shells as well as a  
quantitative description of valence shell electron pair repulsion  
theory~\cite{rjgbook} of bonding 
%(with the notable exception of single 
%covalent bond regions as in the H$_2^+$ and H$_2$ molecules) 
and it has been recently 
extended to excited and time-dependent states~\cite{gross}. Nevertheless,
it has two defects: 
(1) it remains a mathematical construction of dubious physical 
significance~\cite{savin} and (2) it is difficult to calculate ELF beyond~\cite{mss} 
the lowest-order approximation [Eq.~(\ref{belf})] and for this reason it 
can hardly be applied to strongly interacting systems.
%\end{itemize} 

In this paper we propose a more physical way of looking at atomic shells 
and bonds based on the idea of {\it quantum stress focusing}, by which 
we mean the spontaneous formation of  regions of high and low pressure 
in the electron gas.  The existence of such regions is nontrivial 
in view of the fact that the electrostatic potential created by the 
nuclei has no maxima or minima (Earnshaw's theorem).

To understand how pressure maxima 
can arise we must recall a few basic concepts from the mechanics of 
continuum media.  First, we  introduce the stress tensor~\cite{martin,tokatly}, 
$p_{ij}(\rv)$, -- a symmetric rank-2 tensor with the 
property that $p_{ij}$ is the i-th component of the force per unit area 
acting on an infinitesimal surface perpendicular to the j-th axis~\cite{note1}.
The divergence of this tensor, $\sum_j \partial_jp_{ij}(\rv)$ 
($\partial_j$ being the derivative with respect to $r_j$, the $j$-th 
component of $\rv$),  is the $i$-th component of the force per unit volume 
exerted on an infinitesimal element of the electron gas by the surrounding 
electrons.  For convenience, in the following we will exclude
from the stress tensor the electrostatic Hartree contribution,
which is then treated as an external field on equal footing with
the field of the nuclei.

%%%%%%%%%%%%%%%%%%%%%%%%%%%%%%%%%%%%%%%%%%%%%%%%%%%%%%%%%%%%%%%%%%%%%
\begin{figure}[t]
\includegraphics[width=2.5in]{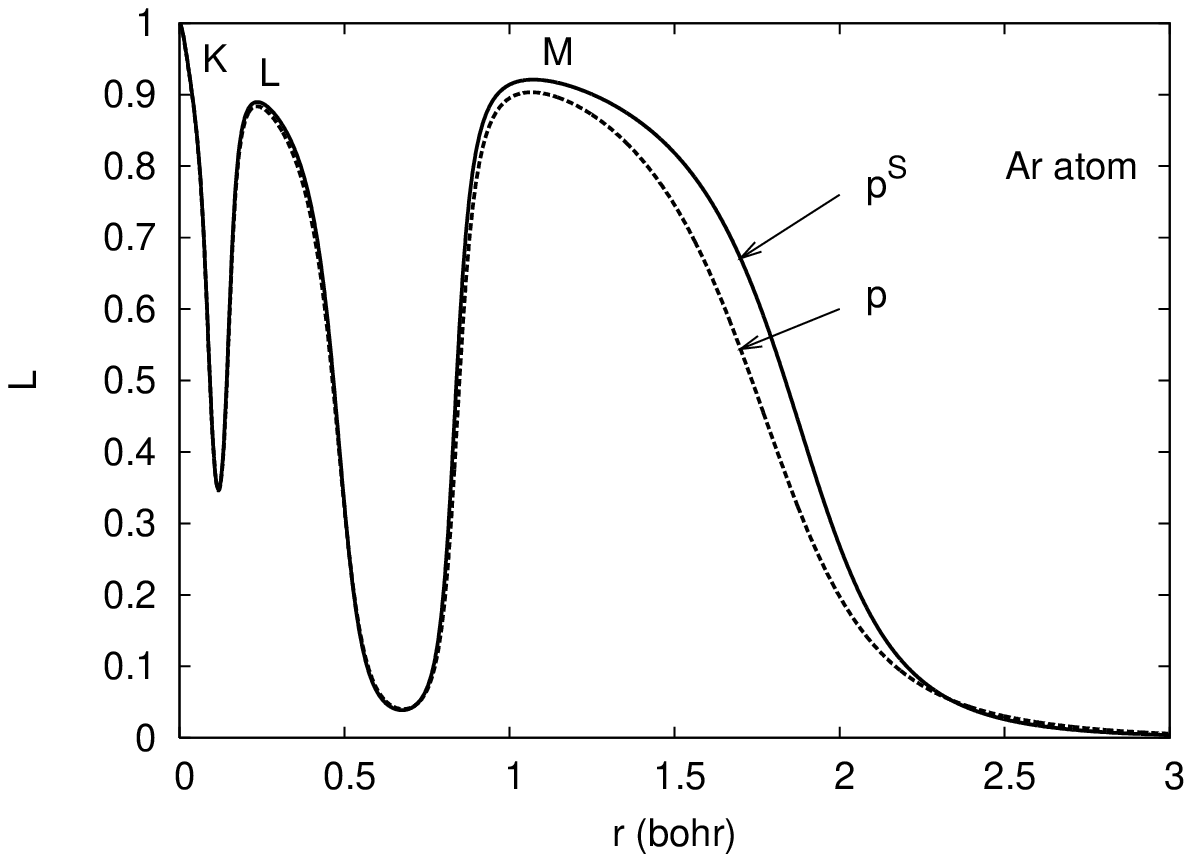}
\includegraphics[width=2.5in]{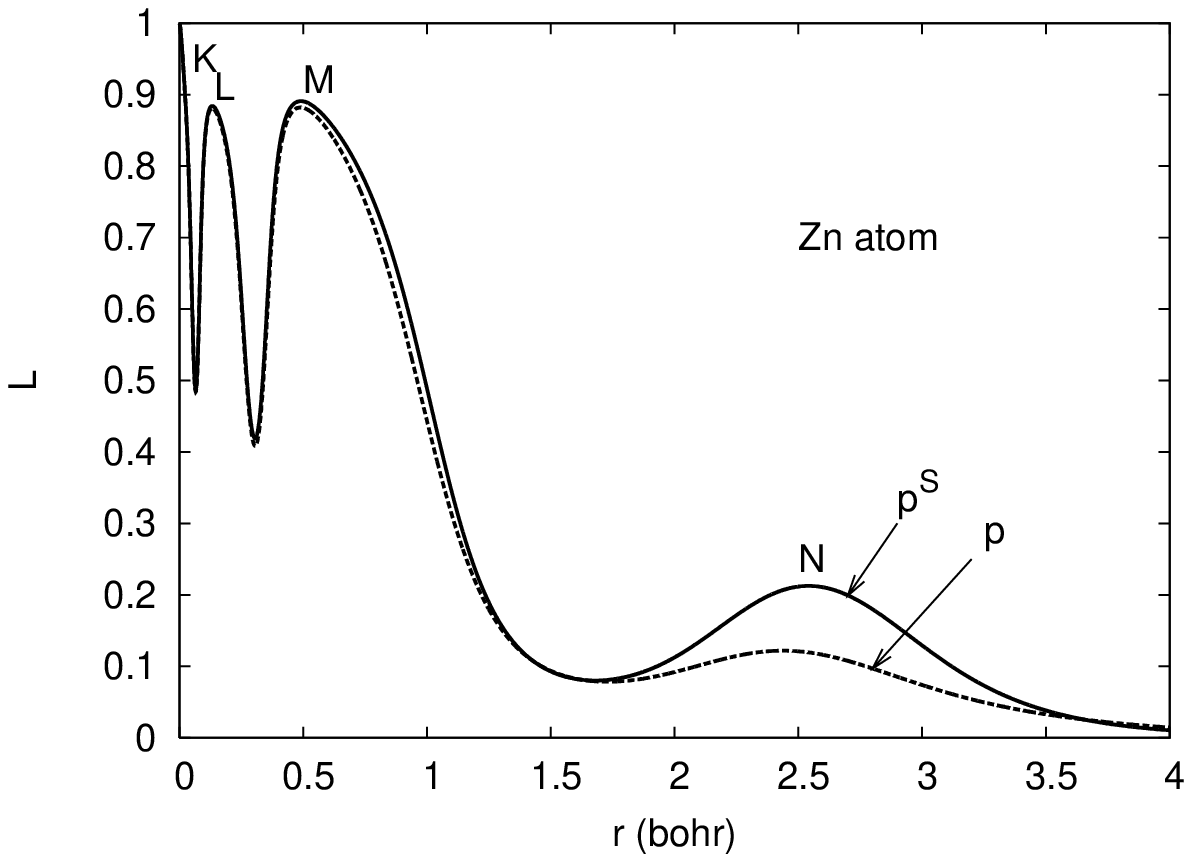}
\end{figure}
\begin{figure}
\includegraphics[width=2.5in]{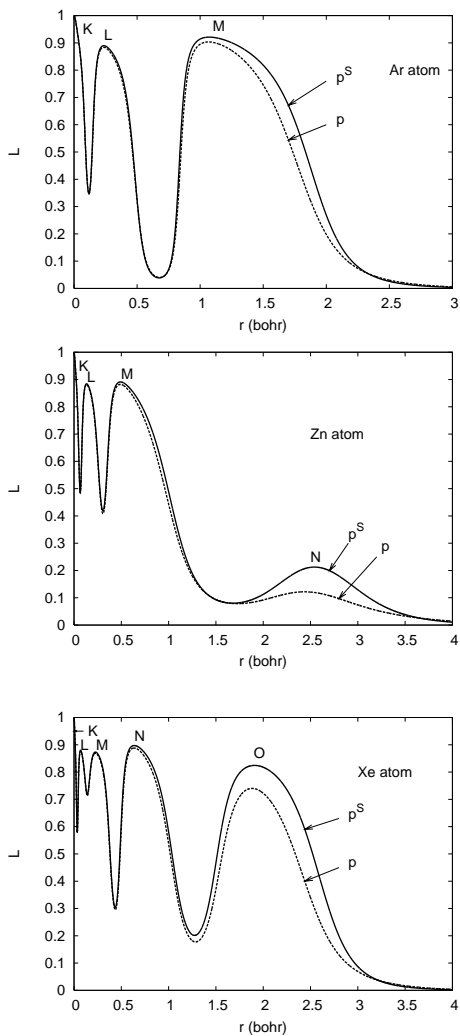}
\caption{Indicators of electron localization  as functions of the
radial distance $r$ constructed respectively
from the non-interacting and interacting pressures $p^{\rm S}$ and $p$
[Eq.~(\ref{indicator})]
for the Ar, Zn, and Xe atoms. The XC part of the pressure is evaluated
with the GGA functional of Ref.~\cite{pbe96}. The maxima of the 
pressure reveal electronic shells, indicated as usual by the capital 
letters K,L,M,...}
\end{figure}
%%%%%%%%%%%%%%%%%%%%%%%%%%%%%%%%%%%%%%%%%%%%%%%%%%%%%%%%%%%%%%%%%%%%%
Next, we separate $p_{ij}$ into a hydrostatic pressure part, 
which is isotropic, and a shear part, which is traceless: 
\begin{eqnarray}\label{split}
p_{ij}(\rv) = 
\delta_{ij}p(\rv) + \pi_{ij}(\rv), \hspace*{0.5cm} {\rm Tr}~\pi_{ij} = 0.
\end{eqnarray}
Here  $p(\rv) \equiv \frac{1}{3}{\rm Tr}~p_{ij}(\rv)$ is the quantum 
pressure, i.e. the derivative of the internal energy with respect to a 
local deformation that changes the volume of a small element of the 
electron liquid without changing its shape.  Similarly, 
the shear stress $\pi_{ij}(\rv)$ is the derivative of the internal 
energy with respect to a local deformation that changes the shape of 
a small element of the liquid, without changing its volume.  In a 
quantum system, these local deformations are implemented through a 
local change in the metrics (see Eq.~(\ref{metrics}) below and 
Ref.~\onlinecite{tokatly} for details). 
In the ground state of an 
electronic system the forces arising from the divergence of $p_{ij}$  
must exactly balance the external force exerted by the nuclei. In other 
words, the stresses satisfy the equilibrium condition
\begin{equation}\label{forcebalance}
-n(\rv) \partial_i v(\rv)=\partial_ip(\rv)+ \sum_j \partial_j\pi_{ij}(\rv),
\end{equation}
where $v(\rv)$ is the external potential due to the nuclei plus the 
Hartree potential. 

Eq.~(\ref{forcebalance}) is not very useful 
if we don't have explicit expressions for $p$ and $\pi_{ij}$.  However, 
certain approximations can be made.  For example,  
replacing the exact quantum pressure by the quasi-classical Fermi pressure 
$p^{\rm TF}(\rv) =  \frac{2}{5} (3 \pi^2)^{2/3} n^{5/3}$ and neglecting the 
shear term yields the well-known Thomas-Fermi equation for the equilibrium 
density.  Because the electrostatic potential $v(\rv)$  in an atom 
does not admit 
maxima and minima, we immediately see that this approximation (as any 
approximation that neglects $\pi$) cannot yield local maxima or minima 
in the pressure. In the special case of the spherical Thomas-Fermi atom 
this implies that the density and the pressure are both monotonically 
decreasing functions of the radial distance $r$:  the shell structure 
is absent. 

The situation changes radically when the shear stress is included.  
Now it is possible for the hydrostatic force $\partial_i p$ to vanish  
because the electrostatic force $n \partial_i  v$ can be balanced by 
the shear force $\sum_j\partial_j\pi_{ij}$. It is therefore possible to 
have maxima or minima in the pressure.  We may picture the regions of 
high pressure as regions which are hard to compress, because of the 
high energy cost of bringing the particles closer together against 
the ``exchange-correlation hole".  These ``maximally compressed 
regions" quite naturally correspond to the shells and bonds of 
descriptive chemistry.   

We now  show that regions of  maximum and minimum pressure do occur 
in spherical atoms, with peaks  corresponding to the K,L,M... shells 
of the standard shell model. To see this we don't need to go any 
further than the lowest-order approximation for the stress tensor, 
which has the form:
\begin{equation}
p_{ij}^{\rm S} = 
\frac{1}{2} \sum_{l\sigma}
\left(\partial_i \psi_{l\sigma}^{*}\partial_j \psi_{l\sigma}
+ \partial_j \psi_{l\sigma}^{*}\partial_i \psi_{l\sigma}\right)
-\frac{1}{4} \delta_{ij}\nabla^2 n,
\end{equation}
where the sum runs over the occupied Kohn-Sham orbitals.
For an atom of spherical symmetry this further simplifies to
\begin{eqnarray}\label{eigentensor}
p_{ij}^{\rm S}= 
p^{\rm S}\delta_{ij}+ \pi^{\rm S}\left(\frac{r_i r_j}{r^2} - 
\frac{\delta_{ij}}{3}\right),
\end{eqnarray}
where $p^{\rm S}$ is the noninteracting pressure given by
\begin{eqnarray}\label{kspressure}
p^{\rm S} = \frac{1}{3}\sum_{l\sigma} |\nabla \psi_{l\sigma}|^2   
- \frac{1}{4}\nabla^2 n,
\end{eqnarray}
and
\begin{eqnarray}\label{Pispressure}
\pi^{\rm S} = \sum_{l\sigma} \left(|\partial_r\psi_{l\sigma}|^2 
- r^{-2}|\partial_\theta\psi_{l\sigma}|^2\right).
\end{eqnarray}

To graphically represent the noninteracting pressure $p^{\rm S}$, we define 
the dimensionless quantity $\tilde p^{\rm S} = p^{\rm S}/p^{\rm TF}$.
Since this ratio diverges both at a nucleus (with a positive sign)  
and in the density tail (with a negative sign)~\cite{note2},
we find it convenient to plot the function 
\begin{eqnarray}\label{indicator}
L(\rv) = \frac{1}{2}\left(1 + \frac{\tilde p^{\rm S}}
{\sqrt{1 + (\tilde p^{\rm S})^2}}\right) 
\end{eqnarray}
which has values in the range $(0,1)$, with $L= 1$ corresponding 
to $\tilde p^{\rm S} =\infty$  and $L= 0$ corresponding to 
$\tilde p^{\rm S} =-\infty$. 
Figures 1a--1c show conclusively the existence of peaks and troughs of 
the noninteracting pressure (solid curves), each of which defines a 
spherical surface on which the gradient of  $p^{\rm S}$ vanishes.  
Since $p^{\rm TF}$ is a monotonically decreasing function, the 
oscillatory behavior of these graphs is entirely due to $p^{\rm S}$: 
in particular, the maxima and minima of $L$ are very close to the 
maxima and minima of $p^{\rm S}$. As discussed above, we interpret the 
peaks in pressure as quantum shells.  Figure 2 shows the behavior of  
$p^{\rm S}$ and  $\pi^{\rm S}$ as functions of $r$, and the inset 
depicts schematically the distribution of shear forces in the atom.   
Observe that the minima of the shear stress are in correspondence with 
the maxima of the pressure, and viceversa. This is natural, because 
the gradients of $p$ and $\pi$, i.e., approximately, the bulk force 
and the shear force must have opposite signs and largely cancel in 
order to balance the relatively weak nuclear attraction. The fact 
that the shell is located at the minimum of the shear stress implies 
that the shear force has opposite signs on the two sides of the 
shell, thus ``locking" the shell into position. 
%The shear stress amplitude $\pi^{\rm S}$  measures the energy cost 
%of increasing (or decreasing) the inner radius of the shell, while 
%simultaneously decreasing (or increasing) its thickness, so that the 
%volume remains unchanged. The maximum  in $\pi^{\rm S}$ can thus be 
%interpreted as an energy barrier defining the boundary between 
%the  K-shell and the L-shell. 
%%%%%%%%%%%%%%%%%%%%%%%%%%%%%%%%%%%%%%%%%%%%%%%%%%%%%%%%
\begin{figure}[t]
\includegraphics[width=2.5in]{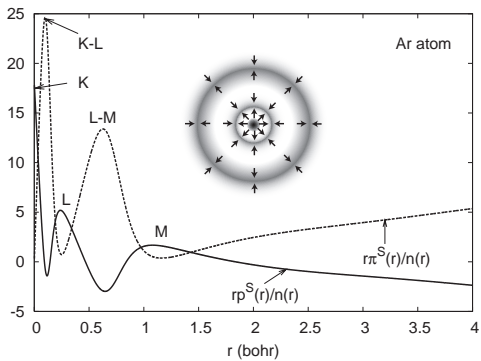}
\caption{Non-interacting pressure ($p^{\rm S}$) and shear stress ($\pi^{\rm S}$) 
pressure in Ar. The 
peaks in $p^{\rm S}$ identify the  position of the shells and  
the peaks in $\pi^{\rm S}$ 
define the boundary between the shells. Inset:  distribution of stresses 
in the atom.  The grey scale indicates the pressure distribution with 
dark (light) regions corresponding to high (low) value of the pressure indicator $L$ (Eq.~\ref{indicator}).  The arrows 
indicate the direction of the shear force $(2/3)\nabla \pi + (2/r)\pi$. 
Observe how the shells are ``squeezed" by shear forces pointing in 
opposite directions.}
\end{figure}
%%%%%%%%%%%%%%%%%%%%%%%%%%%%%%%%%%%%%%%%%%%%%%%%%%%%%%%%%%%%%%

% because the high value of the pressure 
%indicates high electron concentration.  
%The stability of the electron liquid  in the peak region is therefore due 
%entirely to the existence of shear stresses.  
%Similarly, we say that the troughs 
%describe regions of reduced electron concentration in between two shells 
%or intershells (such regions are required by the conservation of the 
%electron number).    

Two more points should be made about Eq.~(\ref{kspressure}) for 
the pressure.  The first is that the Laplacian of the density 
appears in it prominently, as a universal component of the pressure:  
this explains a posteriori the partial success of the Laplacian 
as an indicator of shell structure.  More interestingly, the ELF 
(Eq.~\ref{ELF}) is also closely related to the pressure.
Indeed, the non-interacting pressure includes a ``bosonic'' 
contribution, $p_0$,  which is obtained from Eq.~(\ref{kspressure}) 
by putting all the electrons in the lowest energy orbital 
$\psi_{0 \sigma}$:
%would be present even if all the 
%electrons could be put in the
%same orbital $\psi_{0 \sigma}(\rv) = 
%\sqrt{n_\sigma (\rv)/N_\sigma}$. The bosonic 
%pressure, $p_0$,  can be found from Eq.~(\ref{kspressure}) by putting
%all the electrons in the orbital $\psi_{0 \sigma}$, which gives 
\begin{eqnarray}\label{boson}
p_{0 \sigma} = 
(1/3)[|\nabla n_\sigma|^2/(4n_\sigma) - (3/4)\nabla^2 n_\sigma].
\end{eqnarray}
The excess pressure 
$p_{\rm ex \sigma} \equiv p_{\sigma}^{\rm S} - p_{0 \sigma}(\rv)$ 
is called  ``Pauli pressure'' (since it is generated by the Pauli 
exclusion principle, which forces the occupation of higher energy 
orbitals) and is easily seen to coincide with  $D_\sigma$,  the 
curvature of the conditional probability function (Eq.~(\ref{belf})) 
and main ingredient of the ELF.   
Physically, high electronic pressure means that it is difficult to 
bring the particles closer together, which implies that the XC hole 
surrounding the electron is very ``deep", i.e. the probability of 
finding another electron of the same spin around the reference 
electron is low.  This leads to a high value of the ELF. 
%Physically, high electronic pressure implies that the electron 
%and the XC hole surrounding the electron are both
%highly localized so that it is difficult to bring the particles 
%closer together and thus the probability of finding another electron 
%of the same spin around the reference electron is low, 
%leading to a high value of the ELF. 

%Physically, it is not surprising 
%that high electronic pressure goes hand in hand with a deep 
%exchange-correlation hole, for both features imply that it is 
%difficult to bring the particles closer together. Thus, high pressure 
%implies a small value of $D_\sigma$ and a correspondingly high 
%value of the ELF. 
%Like the Laplacian, we see that the ELF is also part of the quantum 
%pressure.  Again this explains, a posteriori, the success of the ELF as 
%an indicator of electron concentration~\footnote{One must notice however 
%that the Pauli pressure vanishes for just two electrons in a covalent 
%bond: hence the failure of the ELF to indicate the spatial structure 
%of the $H_2$ molecular bond.}. 

One of the attractive features of our proposal is the relative ease 
with which strong electron-electron interactions can be included in 
the calculation of the stress tensor.  Already $p^{\rm S}$ includes 
some interaction effects through the Kohn-Sham orbitals~\cite{ks65}.  
To include additional exchange-correlation (XC) effects we observe 
that in density functional theory (DFT) the XC energy is a functional 
not only of the density $n$, but also, tacitly, of the metrics 
$g_{ij}$ of the space\cite{tokatly}:
%\footnote{\label{metricnote}The metric tensor $g_{ij}$ determines the 
%length $ds$ of an infinitesimal  line element of coordinate increments 
%$dx_i$: $ds^2 = \sum_{ij}g_{ij}dx_idx_j$}: 
$E_{\rm xc} = E_{\rm xc}[n,g_{ij}]$.  

%The above expressions include only non-interacting kinetic contributions.  
%Electron-electron interactions enter only implicitly, via the density and 
%the Kohn-Sham orbitals. The genuine interaction contribution can be 
%calculated with the help of density functional theory (DFT)~  
%Let us  now discuss how to include genuine 
%interaction contributions.  It turns out that this can be efficiently 
%done with the help of density functional theory (DFT).  
%The key observation is that in DFT the exchange-correlation (XC) energy  
%(and the classical Hartree energy as well) is a functional of the density 
%$n$, but also, tacitly, of the metrics 
%$g_{ij}$ of the space\footnote{The metric tensor $g_{ij}$ determines the 
%length $ds$ of an infinitesimal  line element of coordinate increments 
%$dx_i$: $ds^2 = \sum_{ij}g_{ij}dx_idx_j$}: 
%$E_{\rm xc} = E_{\rm xc}[n,g_{ij}]$.  We focus on the XC energy here.  

In conventional applications of DFT, one works with fixed Euclidean metrics 
$g_{ij} = \delta_{ij}$, so there is no need to emphasize the dependence on
the metrics.  Here, however, we are interested in small deformations, 
which can be described by the coordinate transformation~\cite{tvt07} 
$\bm\xi (\rv) = \rv - \uv (\rv)$, where 
$\uv$ is an infinitesimal nonlinear function of $\rv$. The transformation
from $\rv$ to  $\bm \xi$ 
changes the density from $n(\rv)$ to, up to first order in $\uv$,  $\tilde n(\bm\xi ) = n(\bm\xi)+u_i\partial_in(\bm\xi )$  (notice that $i$ here are cartesian indices with the convention that 
repeated indices are summed over), and the metric tensor from the Euclidean value $\delta_{ij}$ to
\begin{eqnarray}\label{metrics}
g_{ij} \equiv (\partial r_\alpha/\partial {\xi_i})
(\partial r_\alpha/\partial {\xi_j}) = \delta_{ij} + \partial_i u_j + 
\partial_j u_i. 
\end{eqnarray}
Since the XC energy cannot depend upon
the arbitrary choice of coordinates we must have 
$E_{\rm xc}[n,\delta_{ij}] = E_{\rm xc}[\tilde n,g_{ij}]$, 
which for an arbitrary infinitesimal transformation implies   
\begin{eqnarray}\label{imply}
n \partial_i \frac{\delta E_{\rm xc}}{\delta n} =
 \partial_j \left\{\delta_{ij}n\frac{\delta E_{\rm xc}}{\delta n} - 2\frac{\delta E_{\rm xc}}{\delta g_{ij}}\right \},
\end{eqnarray}
where the derivative with respect to $n$ is taken at constant $g_{ij}$ and viceversa.
Since the left-hand side of this equation is the XC force density, the right-hand
side must be the divergence of the XC stress tensor. This leads to the 
result
\begin{eqnarray}\label{iden}
p_{ij}^{\rm xc}= \delta_{ij}nv_{\rm xc}-
2 \frac{\delta E_{\rm xc}}{\delta g_{ij}},
\end{eqnarray}
where $v_{\rm xc}$ is the XC potential defined by 
$v_{\rm xc}\equiv \delta E_{\rm xc}/\delta n$, and the functional derivative is evaluated at $g_{ij}=\delta_{ij}$. 
This important result shows that we may calculate the XC contribution to the 
stress tensor from a knowledge of the XC energy as a functional of density 
and metrics. It turns out that semi-local XC functionals~\cite{pbe96,tpss} are 
of a form that can be easily generalized to include a nontrivial metrics.

For example, consider the generalized gradient approximation (GGA) for a 
spin-unpolarized system.
The standard form of this functional is $E_{\rm xc}[n] = \int d \rv 
e_{\rm xc}(n,s)$, where $s= |\nabla n|/(2k_Fn)$ is the reduced density gradient.  
Going to curvilinear coordinates we immediately obtain the generalized form 
\begin{eqnarray}\label{semil}
E_{\rm xc}[n, g_{ij}] = \int d \bm\xi \sqrt{g(\bm\xi)}
e_{\rm xc}(\tilde n,\tilde s),
\end{eqnarray}
where $g \equiv {\rm det}(g_{ij})$ is the determinant of the metric tensor,
$\tilde n (\bm\xi) \equiv n(\rv(\bm\xi))$  is the density in the new coordinates, and 
$\tilde s = \sqrt{g^{ij}\partial_i {\tilde n}\partial_j {\tilde n}}
/(2\tilde k_F \tilde n)$, where $\tilde k_F = (3\pi^2 \tilde n)^{1/3}$
and $g^{ij} = (g^{-1})_{ij}$. 
Using the identity $\delta g = g g^{ij} \delta g_{ij}   = -g g_{ij} \delta g^{ij}$,
we evaluate the functional derivative of $E_{\rm xc}[\tilde n, g_{ij}]$
with respect to $g_{ij}$, and then substitute it into 
Eq.~(\ref{iden}). We finally obtain
\begin{eqnarray}\label{pxcij}
p_{ij}^{\rm xc} = \delta_{ij}(nv_{\rm xc}-e_{\rm xc}) -
\frac{\partial_i n \partial_j n}{2k_Fn|\nabla n|}\frac{\partial e_{\rm xc}}{\partial s}~. 
\end{eqnarray}
The trace of this yields the XC pressure
\begin{eqnarray}\label{xcpressure}
p^{\rm xc} = n v_{\rm xc}-e_{\rm xc} -
\frac{s}{3}\frac{\partial e_{\rm xc}}{\partial s}~, 
\end{eqnarray}
which exactly recovers the uniform-gas limit~\cite{martin}.

In Figs. 1a--1c we have also plotted the normalized pressure $L(r)$ including  
the XC correction (dashed curves), which is calculated by 
the GGA of Ref.~\cite{pbe96}.
In the core region of the atom, where the density 
is high, $p^{\rm xc}$  is utterly negligible. In the valence region, the XC 
contribution becomes relatively important, so the difference is noticeable. 
In the density tail, both $p^{\rm S}$ and $p^{\rm xc}$ 
decay exponentially, causing the difference to be small. 
It is possible, however, and indeed
quite likely, that XC contributions become more important in strongly 
correlated systems, where the zeroth-order description provided by the 
Kohn-Sham orbitals begins to break down. This remains a subject for future 
investigations.

In conclusion, we have found that the analysis of the quantum stress field 
reveals important features of the electronic structure of atoms.  
Although we have considered only atoms so far, there is every reason 
to expect that stress focusing will also occur in molecules in 
correspondence of electron pair domains, such as the ones that are 
commonly associated with single and multiple covalent bonds and with 
the so-called lone-pair regions of the molecule.  In general, the 
local pressure -- the trace of the stress tensor --  is expected to 
provide an excellent quantitative description of electron localization. 
More information may emerge from the complete study of the stress 
tensor (its eigenvalues and principal axes) and its topology -- a 
study that is just beginning to be undertaken.

We acknowledge valuable discussions with C.A. Ullrich and P.L. de Boeij.
This work was supported by DOE under Grant No. DE-FG02-05ER46203, and
by DOE under Contract No. DE-AC52-06NA25396 and Grant No. LDRD-PRD X9KU 
at LANL (J.T).

%%%%%%%%%%%%%%%%%%%%%%%%%%%%%%%%%%%%%%%%%%%%%%

\begin{thebibliography}{100}
\bibitem{note1}
$p_{ij}(\rv)$ is formally defined as the derivative of the internal energy 
with respect to a local strain $u_{ij}(\rv)$, arising from a deformation 
in which each particle is displaced from $\rv_n$ to $\rv_n +\uv(\rv_n)$, 
with $u_{ij} \equiv (\partial_i u_j+\partial_j u_i)/2$. This deformation 
corresponds to a local change in metrics [see Eq.~(\ref{metrics})] which 
modifies the Hamiltonian according to the formulas given in Ref.~\cite{tokatly}.
\bibitem{note2}
The positive divergence at a nucleus is caused by the
Laplacian of the density, which diverges as $1/r$ for $r \to 0$.
The negative divergence in the density tail ($r \to \infty$) reflects
the negative value of $p^{\rm S}$ in the non-classical regions
where the wave function decays exponentially:
$p^{\rm S} \stackrel{r\to \infty} {\to}- 4n|\epsilon_{\rm max}^{\rm KS}|/3$,  
where $\epsilon_{\rm max}^{\rm KS}$ is the highest occupied Kohn-Sham 
orbital energy of the atom.
\end{thebibliography}
\end{document}